\documentclass[11pt,twoside]{article}
\usepackage{newpasp}
\usepackage{epsf}

\index{Matthews, L. D.}
\index{Wood, K.}
\index{Gao, Y.}

\begin{document}

\title{The ISM of Low Surface Brightness Galaxies}
\author{L. D. Matthews}
\affil{NRAO, 520 Edgemont Road, Charlottesville, VA 22903 USA}
\author{Kenneth Wood}
\affil{SAO, 60 Garden Street, Cambridge, MA 02138 USA}
\author{Yu Gao}
\affil{IPAC, MS 100-22, Caltech, Pasadena, CA 91125 USA}

\begin{abstract}
 Using Monte Carlo radiative transfer techniques, we model
the dust content of the edge-on low surface brightness (LSB) galaxy UGC~7321 
and explore the
effects of dust on its disk color gradients. Dust alone
cannot explain the large radial disk color gradients observed in this
galaxy and we find that a
significant fraction ($\sim$50\%) of the dust in UGC~7321 
appears to be contained in a clumped medium, indicating that at least some LSB
galaxies can support a modest multi-phase ISM structure. In addition, we report
some of the first direct detections of molecular gas (CO) in LSB galaxies.
\end{abstract}

\keywords{galaxies: ISM---ISM: dust, extinction---ISM: 
molecules---radiative transfer}

\vspace{-1.0cm}
\section{Introduction}
Low surface brightness (LSB) disk galaxies can be defined as
rotationally-supported galaxies with extrapolated central surface
brightnesses $\ga$1 magnitude 
fainter than the canonical Freeman value of
$\mu_{B,0}$=21.65~mag arcsec$^{-2}$. Although LSB galaxies are found
spanning the full range of Hubble classes, our present work
emphasizes the very latest spiral types (Scd-Sdm).

LSB disk galaxies are now known to be a common product of
galaxy formation, and have received considerable attention during
the past decade (see Impey \& Bothun 1997
for a review). Nonetheless, our knowledge of the
composition and 
conditions of their interstellar media (ISM) is still highly
incomplete.  Such information is  vital
for understanding the evolutionary histories of these galaxies and
how star formation progresses in low-density galaxian environments.

LSB galaxies are typically rich in neutral hydrogen, having
$M_{HI}\sim10^{8}-5\times 10^{9}M_{\odot}$ 
and $M_{HI}/L_{B}\sim$0.5-1.0 
(in solar units). 
A key difference between high surface brightness (HSB) 
and LSB galaxies is that
in LSBs the bulk of the \HI\ often has a surface density
below that needed for efficient star formation (e.g., van der Hulst
{\it et al.} 1993). Nonetheless, the blue colors and modest amounts of
H$\alpha$ emission seen in the majority of LSB galaxies show that
they are still managing to form some new stars.

In spite of the evidence for modest amounts of ongoing
star formation, LSB galaxies are often presumed to devoid of
significant amounts of dust and molecular gas. Lines of 
evidence include:
low metallicities ($\la$1/3Z$_{\odot}$;
e.g.,  R\"onnback \& Bergvall
1995); low statistically-inferred internal extinctions (e.g., Tully
{\it et al.} 1998); highly transparent
disks (e.g., O'Neil {\it et al.} 1998); 
and past failures to detect CO emission (see below). It has also been
argued based on theoretical models that it should be difficult for
molecules to form in LSB galaxies owing to their low densities and the
lack of cooling from metals 
(Mihos {\it et al.} 1999; Gerritsen \& de Blok 1999).

 Unfortunately, directly searching for dust and molecular gas in
LSB galaxies can be challenging. For example, these
galaxies tend to be weak far-infrared and submillimeter sources (e.g.,
Hoeppe {\it et al.} 1994), 
and the presence of dust or dark clouds
is difficult to infer in absorption
against the diffuse and patchy background light from their disks. For
these purposes, {\it edge-on} examples of LSB galaxies thus become particularly
valuable.

\begin{figure}
\vspace{-5.25cm}
\plotfiddle{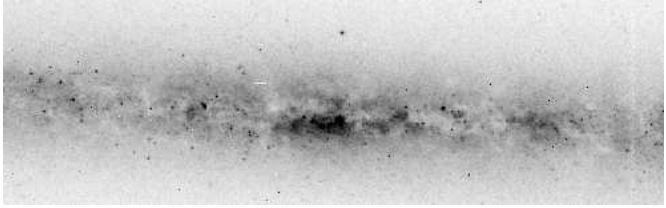}{8.0cm}{0}{100}{100}{-345}{-350}
\caption{\small {\it Hubble Space Telescope} WFPC2 $R+I$ composite
image of the innermost 55$''\times19''$ section of the disk of
UGC~7321. Resolution is $\sim 0.1''\approx$4.8~pc.}
\label{fig1}
\end{figure}

\section{UGC~7321: An Edge-On LSB Galaxy}
Matthews {\it et al.} (1999) have shown that UGC~7321 is an 
example of a nearby ($D\sim$10~Mpc) LSB galaxy viewed near edge-on
($i=88^{\circ}$). 
UGC~7321 has $L_{B}=1.0\times10^{9}~L_{\odot}$, 
$M_{HI}/L_{B}=1.1$($M_{\odot}/L_{\odot}$), and a deprojected central
surface brightness $\mu_{B,i}(0)\sim23.6$~mag~arcsec$^{-2}$. 
A comprehensive description of the  global
properties of this galaxy can be found in
Matthews {\it et al.} (1999).

Fig.~1 shows an 2.6 kpc-long portion of the central disk region of
UGC~7321 as imaged by the WFPC2 on {\it HST}. It is clear from this
image that although UGC~7321 lacks a true dust lane, it is by no means
dust-free. Moreover, the dust is seen to exhibit a clumped structure,
with the size of the individual 
clumps ($\sim$30-60~pc) reminiscent of Giant Molecular Clouds in
the Milky Way.

An intriguing property of UGC~7321 noted by Matthews {\it et al.}
(1999) is that this galaxy exhibits strong radial $B-R$ color
gradients: $\Delta(B-R)\sim 1.0$ from
the disk center to its outermost observed edge. These 
gradients suggest that the disk of UGC~7321 was built up slowly
in time, from the inside out, and that viscous evolution has 
been inefficient. Large radial color gradients have also been seen in
other LSB galaxies (e.g., de Blok {\it et al.} 1995; Bell {\it et al.}
2000; Matthews {\it et al.} 2000). This offers an important clue to the
formation histories of these galaxies. However, properly
interpreting this phenomenon requires accurately correcting for the reddening
effects of dust. 

\section{Monte Carlo Models of the Dust in UGC~7321}
In order to further examine the role of dust in UGC~7321, we
have computed
3-D Monte Carlo radiative transfer models of UGC~7321 at
optical and NIR wavelengths over a range of optical
depths. Our models were computed
on a 200$^{3}$ element grid (resolution $\sim$53~pc) 
and included both
absorption and multiple scattering of photons. We 
employed ``forced first scattering'' and a ``peeling-off'' procedure in
order to allow us to efficiently
treat the very low optical depth regimes encountered in
LSB galaxies.  Our models assume exponential stellar and dust distributions and
Galactic dust grain properties. 
These simulations are described in detail in
Matthews \& Wood (2000). Here we highlight a few key results.

One output from our Monte Carlo models is scattered light images. From
these we find
that models containing only a smoothly distributed dust
component are unable to reproduce the observed morphology of
UGC~7321. Purely smooth models, even with very low optical depths,
exhibit a dust lane, a pronounced central concentration of light,
and a clear asymmetry about the disk midplane not seen in the real
galaxy. 
We instead find the best-fitting models are those with a significant
fraction ($\sim$50\%) of the dust in a {\it clumpy} medium.  This implies 
that interstellar pressures are
sufficient to sustain at least a modest multi-phase ISM in some 
LSB galaxies. Our best 2-phase ISM model for UGC~7321 has a mean edge-on
$B$-band optical depth toward the galaxy center
$\bar\tau_{B,e}\sim$4.0, equivalent to a total dust mass 
$M_{d}\sim8.8\times 10^{5}~M_{\odot}$ and an extinction toward the
galaxy center $A_{B}\sim$0.65 magnitudes. Within the central regions
of the stellar disk, the gas-to-dust ratio appears to be at least several
times the Galactic value, while the global gas-to-dust ratio for
UGC~7321 is $\sim$1600.

From our Monte Carlo models we are able to rule out that the
large radial color gradient observed in UGC~7321 can be explained
fully by dust. In fact, our  models show that for
$\tau_{B,e}\ge4.0$, the
dust-induced $B-R$ radial color gradient {\it saturates} at
$\Delta(B-R)\sim$0.31---i.e. no additional amount of dust can further
increase this gradient (Fig.~2). This establishes that the large radial color
gradients observed in UGC~7321 and other similar late-type LSB
galaxies must instead arise from significant
stellar population and/or metallicity gradients.

\vspace{-0.5cm}
\section{Deep CO Observations of LSB Galaxies}
Since LSB galaxies are actively star-forming systems that can harbor dust and
dark nebulae, it seems likely that  molecular gas should be present in
these galaxies. However, it is unclear how well the CO molecule might be
expected to trace H$_{2}$ in the 
diffuse, low-metallicity environments of LSB
galaxies, and
past searches have failed to detect CO emission from such
objects (Schombert {\it et al.}
1990; Knezek 1993; de Blok \& van der Hulst 1998). 
Nonetheless, late-type spirals are in general not strong CO
emitters (cf. Young \& Knezek 1989), and 
past CO searches in LSB galaxies have employed only modest
integration times.

Using the NRAO 12-m telescope we have obtained very deep, high-quality CO(1-0)
observations of a sample of 6 edge-on LSB galaxies, with total
integration times $\ga$15 
hours per object toward their optical centers. Weather conditions were
excellent, and typically $T_{sys}\sim$300~K. The details of this
study will be described further in Matthews \& Gao (in prep.).

As a result of our deep observations, we have detected 3 of
the 6 LSB galaxies at a $>4\sigma$
level, and one additional source was marginally detected ($\sim 3
\sigma$). Our CO(1-0)
spectrum of the edge-on LSB 
UGC~7321 is shown in Fig.~3. In all cases, the detected CO
lines are relatively broad
($W_{20}\sim$200~km s$^{-1}$) and similar to the global
\HI\ linewidths, suggesting the CO in our targets is spread over a regions
a few kpc across, as is commonly seen in normal very late-type spirals.

Assuming a standard Galactic  conversion factor, the
H$_{2}$ masses inferred for the detected galaxies are $\sim1-5\times
10^{7}~M_{\odot}$, with $M_{H2}/M_{HI}\sim$0.02-0.03. In comparison,
typical $M_{H2}/M_{HI}$ ratios for HSB Sd galaxies are 0.19$\pm$0.10
(Young \& Knezek 1989), implying that LSB galaxies are
comparatively molecular gas-deficient. However, we caution that the 
the validity of the Galactic
CO-to-H$_{2}$ factor for LSB systems is still a matter of debate, and
will require further investigation.

\begin{figure}
\vbox {
  \begin{minipage}[l]{0.7\textwidth}
   \hbox{
       \begin{minipage}[l]{0.5\textwidth}
       {\centering \leavevmode \epsfxsize=\textwidth 
        \epsfbox{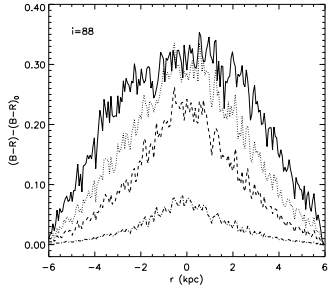}}
       \end{minipage} \  \hfill \
  }
  \end{minipage} \  \hfill \
  \begin{minipage}[c]{0.5\textwidth}
\caption{\small Dust-induced 
$B-R$ color gradients along the galaxy major axis computed
from our clumpy, 2-phase 
Monte Carlo models for four different edge-on optical depths:
$\bar\tau_{B,e}$=0.4, 2.0, 4.0, \& 8.0 and a viewing angle $i=88^{\circ}$.}
  \end{minipage}
}
\label{fig2}
\end{figure}

\begin{figure}
\vspace*{-2.6cm}
\vbox {
  \begin{minipage}[r]{0.50\textwidth}
   \hbox{
       \begin{minipage}[r]{0.8\textwidth}
       {\centering \leavevmode \hspace{+1.5cm} \epsfxsize=\textwidth 
        \epsfbox{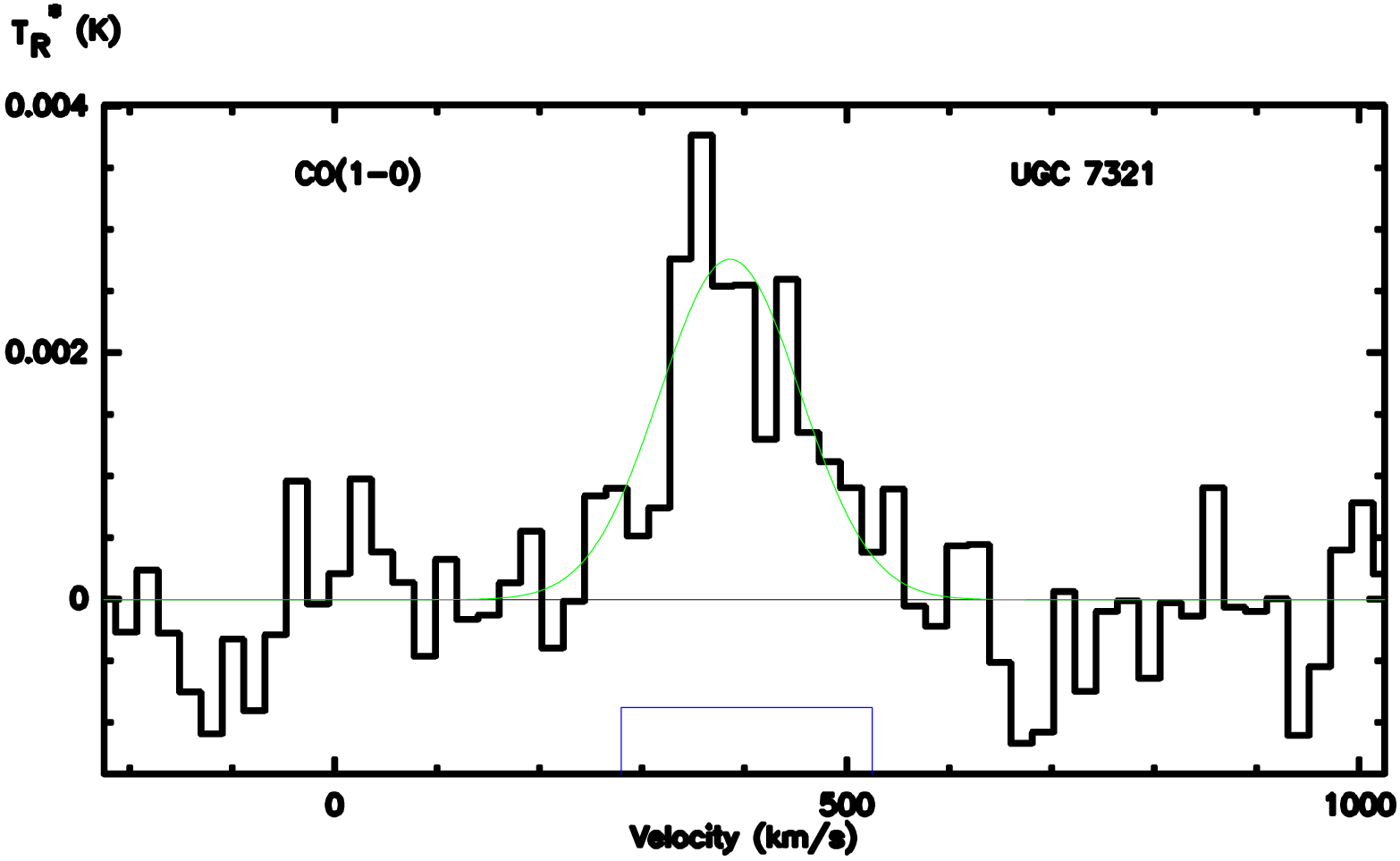}}
       \end{minipage} \  \hfill \
  }
  \end{minipage} \  \hfill \
  \begin{minipage}[c]{0.5\textwidth}
\caption{\small  CO(1-0) spectrum of the central region of 
UGC~7321 obtained with the NRAO
12-m telescope. Total integration time was $\sim$15.2 hours. The
data were smoothed to a resolution of
$\sim$21~km~s$^{-1}$. The beam size was 55$''$. }
  \end{minipage}
}
\label{fig2}
\vspace*{-1.4cm}
\end{figure}

\end{document}